\documentclass{cpbtex}
\usepackage{graphicx}
\usepackage{amsmath}
\usepackage{xcolor}
\newcommand{\BSS}[1]{{\color{black}#1}}
\begin{document}
\begin{CJK*}{GBK}{song}
\title{Anelasticity to plasticity transition in a model two-dimensional amorphous solid \thanks{Project supported by Guangdong Major Project of Basic and Applied Basic Research, China (Grant No.2019B030302010),the NSF of China (Grant No.52130108), Guangdong Basic and Applied Basic Research, China (Grant No.2021B1515140005), Pearl River Talent Recruitment Program (Grant No.2021QN02C04).}
}


\author{Shang Baoshuang$^{1}$, \thanks{Corresponding author. E-mail:shangbaoshuang@sslab.org.cn}\\
$^{1}${Songshan Lake Materials Laboratory, Dongguan 523808, China}\\  
}

\date{\today}
\maketitle

\begin{abstract}
Anelasticity, as an intrinsic property of amorphous solids, plays a significant role in understanding their relaxation and deformation mechanism. However, due to the lack of long-range order in amorphous solids, the structural origin of anelasticity and its distinction from plasticity remain elusive. In this work, \BSS{using frozen matrix method,} 
we study the transition from anelasticity to plasticity transition in a two-dimensional model glass.
Three distinct mechanical behaviours, namely elasticity, anelasticity, and plasticity, are identified with control parameters in the amorphous solid. Through the study of finite size effects on these mechanical behaviors, it is revealed that anelasticity can be distinguished from plasticity. Anelasticity serves as an intrinsic bridge connecting the elasticity and plasticity of amorphous solids. Additionally, it is observed that anelastic events are localized, while plastic events are subextensive. The transition from anelasticity to plasticity is found to resemble the entanglement of long-range interactions between element excitations. This study sheds light on the fundamental nature of anelasticity as a key property of element excitations in amorphous solids.
\end{abstract}



\section{Introduction}

Amorphous solids, as non-equilibrium materials, have captivated the interest of both academic researchers and industrial applications due to their diverse deformation behaviors \ucite{Schuh20074067,Wang2012a,hufnagel2016deformation}. However, unlike crystalline materials, the underlying deformation mechanism in amorphous solids remains a topic of debate. While crystalline materials exhibit topological defects such as dislocations or grain boundaries as \BSS{plasticity carriers}, amorphous solids rely on shear transformations or rearrangement events rather than specific topological structures\ucite{Cheng2011379,barrat2011heterogeneities,RevModPhys.90.045006}.

Previous studies\ucite{argon1979plastic,PhysRevLett.95.095502,schall2007structural,PhysRevE.82.055103,PhysRevE.96.033002}  have observed that rearrangement events in amorphous solids are localized and \BSS{can} occur within the apparent elastic regime. However, considering plastic rearrangements as excitation elements is inappropriate due to their entanglement with long-range elastic interactions and plastic events\ucite{PhysRevE.79.066109,PhysRevLett.93.016001,krisponeit2014crossover,PhysRevLett.112.155501,Lagogianni2018,Shang2019,PhysRevMaterials.7.013601}. Consequently, identifying the excitation element in amorphous solids remains a topic of ongoing debate\ucite{barrat2011heterogeneities,argon2013strain}.

Furthermore, it is crucial to note that not all rearrangement events in amorphous solids exhibit irreversible plastic behavior. Recently, a new type of rearrangement event characterized by an intrinsic anelastic nature has garnered significant attention\ucite{xu2017strain,PhysRevMaterials.4.113609,PhysRevE.102.033006,Regev2015}. Unlike plastic events, these anelastic rearrangements are reversible during the loading-unloading process. They play a vital role in influencing various mechanical properties of amorphous solids, including relaxation \ucite{PhysRevLett.99.135502,Wang2019}, thermal cycling rejuvenation\ucite{Costa2022anelastic}, mechanical anisotropy\ucite{PhysRevB.48.3048,Shang2022cycle} , and memory effect\ucite{PhysRevE.88.062401,PhysRevLett.112.025702}.

However, the understanding of the discrepancies and connections between anelastic and plastic events, as well as the key parameters governing the transition from anelasticity to plasticity, remain elusive. Further investigations are necessary to unravel the underlying mechanisms and establish a comprehensive understanding of the relationship between these two deformation modes in amorphous solids.

In this study, we aim to address these questions through molecular dynamics simulations. We investigate the mechanical response of amorphous solids using athermal quasistatic shear and frozen matrix methods. Our focus is on observing the transition from anelasticity to plasticity and understanding the underlying mechanisms. By analyzing the characterized parameters, we identify three distinct deformation modes: elasticity, anelasticity, and plasticity. Specifically, we explore the effects of finite system size on these parameters. Our findings reveal that anelasticity precedes plasticity and serves as a critical intermediary between elasticity and plasticity. Furthermore, we observe that anelastic events, characterized by their system size independence, exhibit a localized nature within the material. On the other hand, plastic events display subextensive behavior. This suggests that plasticity emerges from the collective interaction of a series of anelastic events. As a result, anelastic events can be regarded as potential element excitations in amorphous solids.

\section{Method}
\subsection{Initial sample preparation}

We used a well studied two dimensional binary \BSS{Lenard-Jones} model \BSS{with a polymoinal cutoff smooth} \ucite{PhysRevE.97.033001,PhysRevMaterials.4.113609} to investigate the mechanical property of amorphous solid. All units were expressed in terms of the mass $m$ and
the two parameters describing the energy and length scales of interspecies interaction, $\epsilon$ and $\sigma$ and the Boltzmann constant $k_B=1$, respectively.
Therefore, time was measured in units of $t_0 = \sqrt{m\sigma^2/\epsilon}$, and temperature was measured in units of $T_0=\epsilon/k_B$. 
The composition ratio of large ($L$) and small ($S$) was $N_L : N_S = (1 + \sqrt{5})/4$.  100 samples each containing $10^4$ atoms with periodic boundary condition, were obtained by quenching from $T=2.0$ to $T=0.18$ with constant volume, and quench rate was fixed at 0.0000325 $T_0/t_0$, and then the initial sample was obtained by minimized the quenched sample with  conjugate gradient (CG) method, where the \BSS{fictive} temperature of the sample is around 0.335.  The reduce density of system was fixed at 1.02. 
All the results were presented in reduced units.
 We performed all the simulations using the LAMMPS molecular dynamics package\ucite{Thompson2022}, and used the OVITO package\ucite{Stukowski2009VisualizationAA} for atomic visualization.

\subsection{Frozen matrix method}

Frozen matrix method is a useful tool to investigate the local yield stress\ucite{PhysRevLett.117.045501} or local modulus\ucite{PhysRevE.87.042306} of the amorphous system, and local relaxation time\ucite{Shang2019} of supercooled liquid, \BSS{it can also use to study the thermaly activated rearrangement\ucite{PhysRevLett.129.195501} and to calibrate more quantitatively elastoplastic models\ucite{FernndezCastellanos2021}, even for different preparation protocol\ucite{Castellanos2022}}.
Here, we extent this method to study the anelasticity to plasticity transition in the amorphous solid.
The investigated region was selected within a radius $R$, then the outside region was frozen with affine deformation (Figure \ref{fig:1} (a)). 
\BSS{
In this study, the investigated region is fixed at the center of each sample.
}
The  mechanical response of amorphous solid was probed by the athermal quasi-static shear (AQS) protocol. 
The simple shear deformation was performed in all regions and the shear strain increment was $\delta \gamma=10^{-5}$, during each shear strain increment, the investigated region was relaxed by energy minimization, and the outside region was frozen with affine deformation. 
The loading process continues until the shear strain reaches to 0.3. \BSS{
The choice of 0.3 is not grounded in specific physical reasoning; rather, it was selected to ensure that all investigated samples exhibit a transition from anelasticity to plasticity.
}
\subsection{Physical property characterization}
\subsubsection{anelastic event and plastic event}
During the loading process, stress and potential energy drops occur (Figure \ref{fig:1} (b),(c)), which are caused by rearrangement events (Figure \ref{fig:1} (d)). A rearrangement event is defined as when the maximum atomic displacement exceeds 0.1, or the potential energy drop per atom $\Delta U$ is greater than 0.1, in the investigated region. To characterize the anelastic and plastic events, the following unloading process was performed. After each potential energy drop, the shear strain was reversed and sheared back to $\gamma = 0$, which is called the unloaded sample. The mean squared displacement (MSD) between the unloaded sample and the initial sample was compared, and if MSD was zero, the drop was an anelastic event; otherwise, it was a plastic event. The potential energy per atom $U$ and the shear stress $\tau$ of the investigated region were monitored during the loading and unloading process, and the shear strain increment was fixed at $\delta \gamma=10^{-5}$.
\subsubsection{atomic displacement}
To compare configuration 1 and configuration 2, the atomic displacement of atom $i$ is defined as $\vec{d_i}=\vec{r_i}(1)-\vec{r_i}(2)$, where $\vec{r_i}(1)$ is the coordination vector of atomic $i$ at configuration 1. The MSD between two configuration can be defined as $1/N \sum_{i \in N} \vec{d_i} \cdot \vec{d_i}$, where $N$ is the atomic number in the investigation region. 

\section{Method}
\subsection{Initial sample preparation}

We used a well studied two dimensional binary \BSS{Lenard-Jones} model \BSS{with a polymoinal cutoff smooth} \ucite{PhysRevE.97.033001,PhysRevMaterials.4.113609} to investigate the mechanical property of amorphous solid. All units were expressed in terms of the mass $m$ and
the two parameters describing the energy and length scales of interspecies interaction, $\epsilon$ and $\sigma$ and the Boltzmann constant $k_B=1$, respectively.
Therefore, time was measured in units of $t_0 = \sqrt{m\sigma^2/\epsilon}$, and temperature was measured in units of $T_0=\epsilon/k_B$. 
The composition ratio of large ($L$) and small ($S$) was $N_L : N_S = (1 + \sqrt{5})/4$.  100 samples each containing $10^4$ atoms with periodic boundary condition, were obtained by quenching from $T=2.0$ to $T=0.18$ with constant volume, and quench rate was fixed at 0.0000325 $T_0/t_0$, and then the initial sample was obtained by minimized the quenched sample with  conjugate gradient (CG) method, where the \BSS{fictive} temperature of the sample is around 0.335.  The reduce density of system was fixed at 1.02. 
All the results were presented in reduced units.
 We performed all the simulations using the LAMMPS molecular dynamics package\ucite{Thompson2022}, and used the OVITO package\ucite{Stukowski2009VisualizationAA} for atomic visualization.

\subsection{Frozen matrix method}

Frozen matrix method is a useful tool to investigate the local yield stress\ucite{PhysRevLett.117.045501} or local modulus\ucite{PhysRevE.87.042306} of the amorphous system, and local relaxation time\ucite{Shang2019} of supercooled liquid, \BSS{it can also use to study the thermaly activated rearrangement\ucite{PhysRevLett.129.195501} and to calibrate more quantitatively elastoplastic models\ucite{FernndezCastellanos2021}, even for different preparation protocol\ucite{Castellanos2022}}.
Here, we extent this method to study the anelasticity to plasticity transition in the amorphous solid.
The investigated region was selected within a radius $R$, then the outside region was frozen with affine deformation (Figure \ref{fig:1} (a)). 
\BSS{
In this study, the investigated region is fixed at the center of each sample.
}
The  mechanical response of amorphous solid was probed by the athermal quasi-static shear (AQS) protocol. 
The simple shear deformation was performed in all regions and the shear strain increment was $\delta \gamma=10^{-5}$, during each shear strain increment, the investigated region was relaxed by energy minimization, and the outside region was frozen with affine deformation. 
The loading process continues until the shear strain reaches to 0.3. \BSS{
The choice of 0.3 is not grounded in specific physical reasoning; rather, it was selected to ensure that all investigated samples exhibit a transition from anelasticity to plasticity.
}
\subsection{Physical property characterization}
\subsubsection{anelastic event and plastic event}
During the loading process, stress and potential energy drops occur (Figure \ref{fig:1} (b),(c)), which are caused by rearrangement events (Figure \ref{fig:1} (d)). A rearrangement event is defined as when the maximum atomic displacement exceeds 0.1, or the potential energy drop per atom $\Delta U$ is greater than 0.1, in the investigated region. To characterize the anelastic and plastic events, the following unloading process was performed. After each potential energy drop, the shear strain was reversed and sheared back to $\gamma = 0$, which is called the unloaded sample. The mean squared displacement (MSD) between the unloaded sample and the initial sample was compared, and if MSD was zero, the drop was an anelastic event; otherwise, it was a plastic event. The potential energy per atom $U$ and the shear stress $\tau$ of the investigated region were monitored during the loading and unloading process, and the shear strain increment was fixed at $\delta \gamma=10^{-5}$.
\subsubsection{atomic displacement}
To compare configuration 1 and configuration 2, the atomic displacement of atom $i$ is defined as $\vec{d_i}=\vec{r_i}(1)-\vec{r_i}(2)$, where $\vec{r_i}(1)$ is the coordination vector of atomic $i$ at configuration 1. The MSD between two configuration can be defined as $1/N \sum_{i \in N} \vec{d_i} \cdot \vec{d_i}$, where $N$ is the atomic number in the investigation region. 

\section{Results and Discussions}
\begin{figure}[!htbp]
    \centering
    \includegraphics[width=0.8\textwidth]{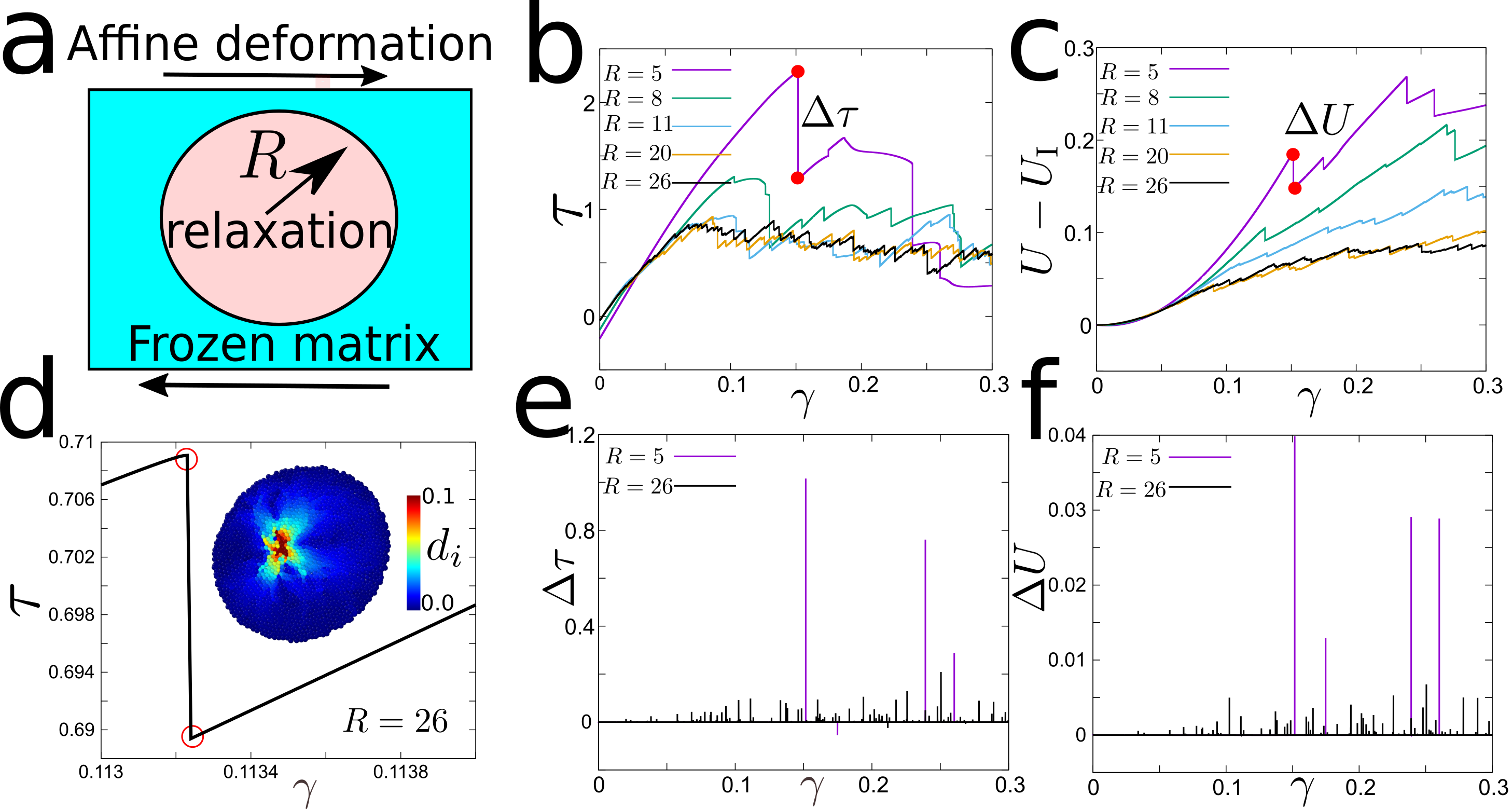}
    \caption{\textbf{The loading process of the amorphous solid.} \textbf{(a)} the schematic of frozen matrix deformation. 
    \textbf{(b)} the shear stress $\tau$ changes with shear strain $\gamma$ in various systems.
    \textbf{(c)} the potential energy $U-U_\text{I}$ changes with strain in various systems.
    \textbf{(d)} the typical stress drops in $R=26$ system and the corresponding atomic displacement color map shown in the panel. 
    \textbf{(e)} the stress drops ($\Delta \tau$) changes with strain in $R=5$ and $R=26$ systems.
    \textbf{(f)} the potential energy drop ($\Delta U$) changes with strain in $R=5$ and $R=26$ systems, respectively.
    }
    \label{fig:1}
\end{figure}
\begin{figure}[!htpb]
    \centering
    \includegraphics[width=0.8\textwidth]{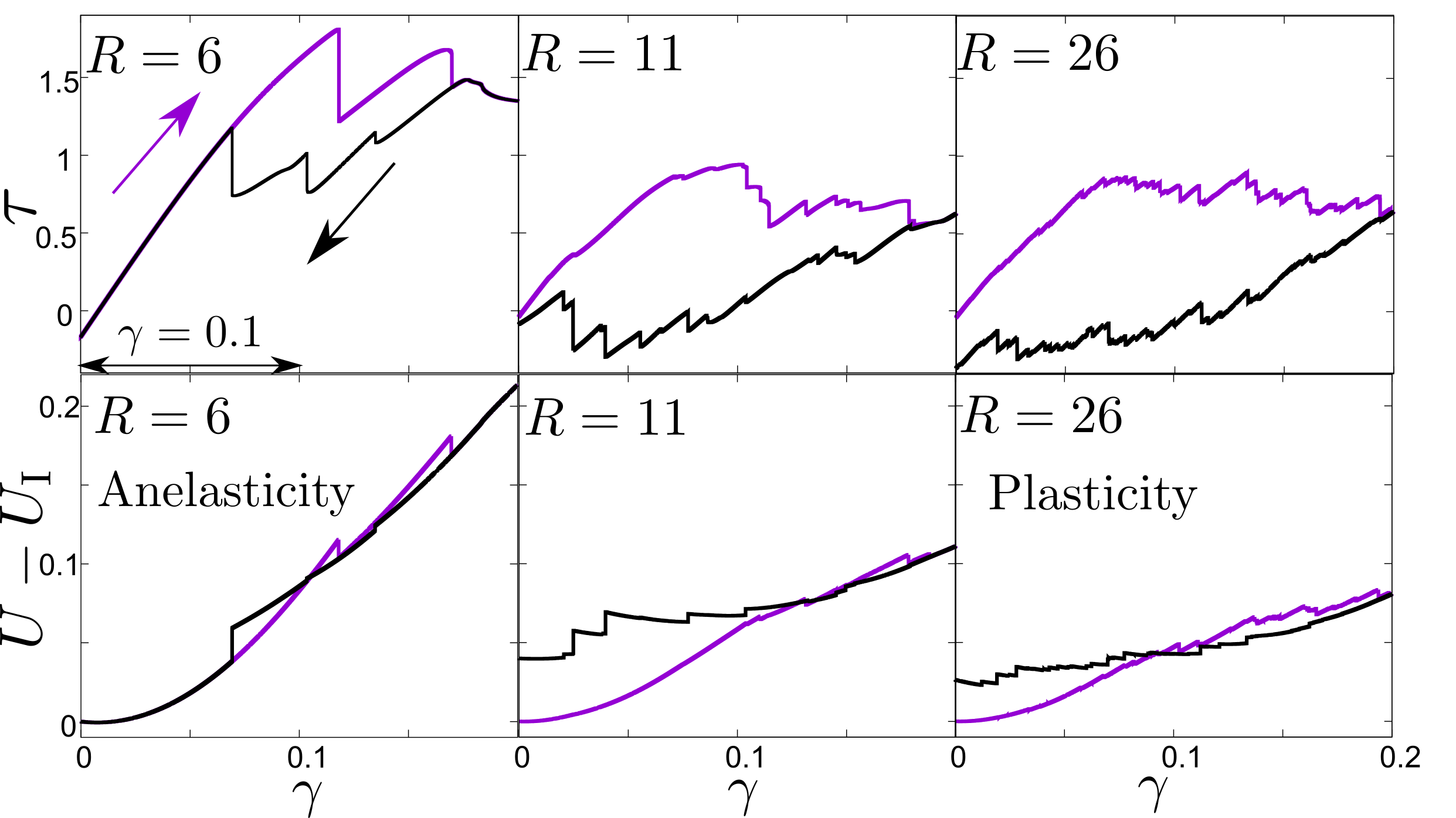}
    \caption{\textbf{The loading and unloading process for various system sizes}. Up panel shows the shear stress involves with shear strain during loading and unloading process, down panel shows the potential energy involves with shear strain.  The purple arrow and black arrow  illustrates the loading and unloading directions, respectively.}
    \label{fig:2}
\end{figure}

Figure \ref{fig:1} shows the loading process of the amorphous solid with various system sizes. 
There are notable finite size effects with frozen matrix boundary condition, both the \BSS{frequency} of stress drop $\Delta \tau$ and potential energy drop per atom $\Delta U$ increases with system size $R$. Conversely, the \BSS{amplitude} of $\Delta \tau$ and $\Delta U$ decreases with system size (Figure \ref{fig:1} \BSS{e,f}). 
During the stress drop (Figure \ref{fig:1} e), the atomic displacement $d_i$ shows a typical \BSS{quadrupolar} symmetry. 
This is qualitatively consistent with the situation with periodic boundary condition \ucite{lemaitre2006sum,PhysRevE.82.055103}.
However, the frozen matrix boundary condition not only blocks the long range elastic interaction from outside region, 
but also causes significant confinement effect\ucite{Shang2019}. 
As suggested by Regev et al \ucite{Regev2015}, the confinement effect can lead to reversible rearrangement. As shown in Figure \ref{fig:2} , for $R=6$ system, both the stress and potential energy state are fully recovered after the loading-unloading process, indicating an anelastic behavior.
In contrast, for $R=26$ system, both the stress and potential energy states are different from the initial state, indicating a plastic behavior. 
Interestingly, for $R=11$ system, the stress state is almost recovered but the energy state is not. 
This behavior can be attributed to the confinement effect of the frozen matrix boundary, and it signifies a transition from anelasticity to plasticity controlled by the system size. Increasing the system size weakens the confinement effect, thereby influencing the nature of the deformation response.

\begin{figure}[!htpb]
    \centering
    \includegraphics[width=0.8\textwidth]{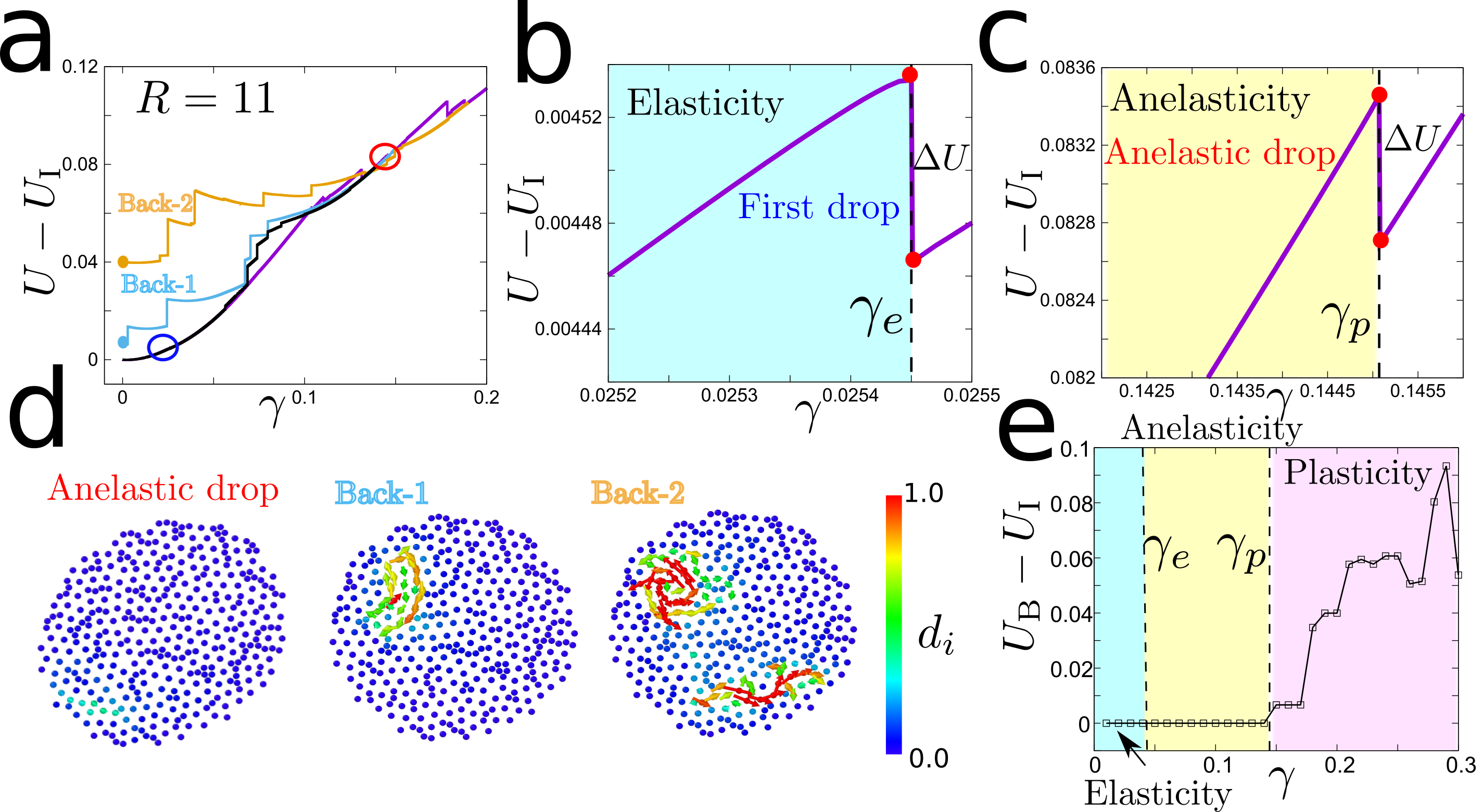}
\caption{
\textbf{The characterization of elastic, anelastic and plastic events by loading-unloading process.} (a) The potential energy per atom ($U$) changes with strain during loading and unloading processes, where $U_I$ is the potential energy per atom of the initial state. The purple curve shows the loading process, and the black one shows the anelastic unloading process, starting from a strain of $\gamma = 0.14$. The yellow and cyan unloading curves start from $\gamma = 0.15$ and 0.19, respectively, and Back-1 and Back-2 indicate the corresponding unloaded states. (b) The first drop event during the loading process, marked by a blue circle in figure (a), and the strain of the first drop event ($\gamma_e$), below which the unloading curve matches the loading curve. (c) The last anelastic drop event during the loading process, and the corresponding strain ($\gamma_p$), marked by a red circle in figure (a).
\BSS{(d) Depiction of atomic displacement in three sample scenarios: anelastic drops, back-1, and back-2. (e) Evolution of potential energy difference from the unloaded state to the initial state with loading strain. Three distinct regions are illustrated by $\gamma_e$ and $\gamma_p$, indicated by the vertical dash line.} 
}
    \label{fig:3}
\end{figure}
Moreover, it should be noted that for a given system size, the increase in strain can also lead to a transition from anelasticity to plasticity. This behavior is depicted in Figure \ref{fig:3}, which illustrates the loading process with various unloading processes for a system size of $R = 11$. 
When the unloading process starts from a strain value of $\gamma = 0.14$, the potential energy of the system can be fully recovered to its initial state. 
However, when the unloading process starts from $\gamma = 0.15$ or $\gamma = 0.19$, the potential energy of the unloaded samples (Back-1, Back-2) is higher than the initial state (Figure \ref{fig:3}(a)).
The loading process exhibits a characteristic first drop strain, denoted as $\gamma_e = 0.02545$ (Figure 3(b)). When the loading strain $\gamma$ is smaller than $\gamma_e$, the loading-unloading process can be fully recovered. This behavior is indicative of elasticity, where no dissipation ($\oint \tau d\gamma \equiv 0$) occurs during the loading phase.
In addition, when the loading strain exceeds $\gamma_e (\gamma > \gamma_e)$, the potential energy-strain curve exhibits sudden drops, indicating atomic rearrangement events. These energy drops are further examined through the unloading process. The last energy drop corresponding to an anelastic event can be identified (Figure \ref{fig:3}(c)). The loading process exhibits a last anelastic drop strain denoted as $\gamma_p = 0.145$. For loading strains between $\gamma_e$ and $\gamma_p$, the loading-unloading process can be fully recovered, but dissipation occurs. When the loading strain exceeds $\gamma_p (\gamma > \gamma_p)$, the unloaded sample cannot be fully recovered to its initial state, and plastic rearrangement takes place.
The displacement color map of the anelastic drop is presented in Figure \ref{fig:3}(d). A comparison of the atomic displacement during the anelastic drop with the plastic displacement of Back-1 and Back-2 samples reveals that the atomic displacement during the anelastic drop is reversible, while plastic rearrangement results in permanent displacement. By comparing the energy of the unloaded sample with the initial state, three types of mechanical property during the loading process can be identified: elasticity, anelasticity, and plasticity (Figure \ref{fig:3}(e)).
\begin{figure}[!htpb]
    \centering
    \includegraphics[width=0.8\textwidth]{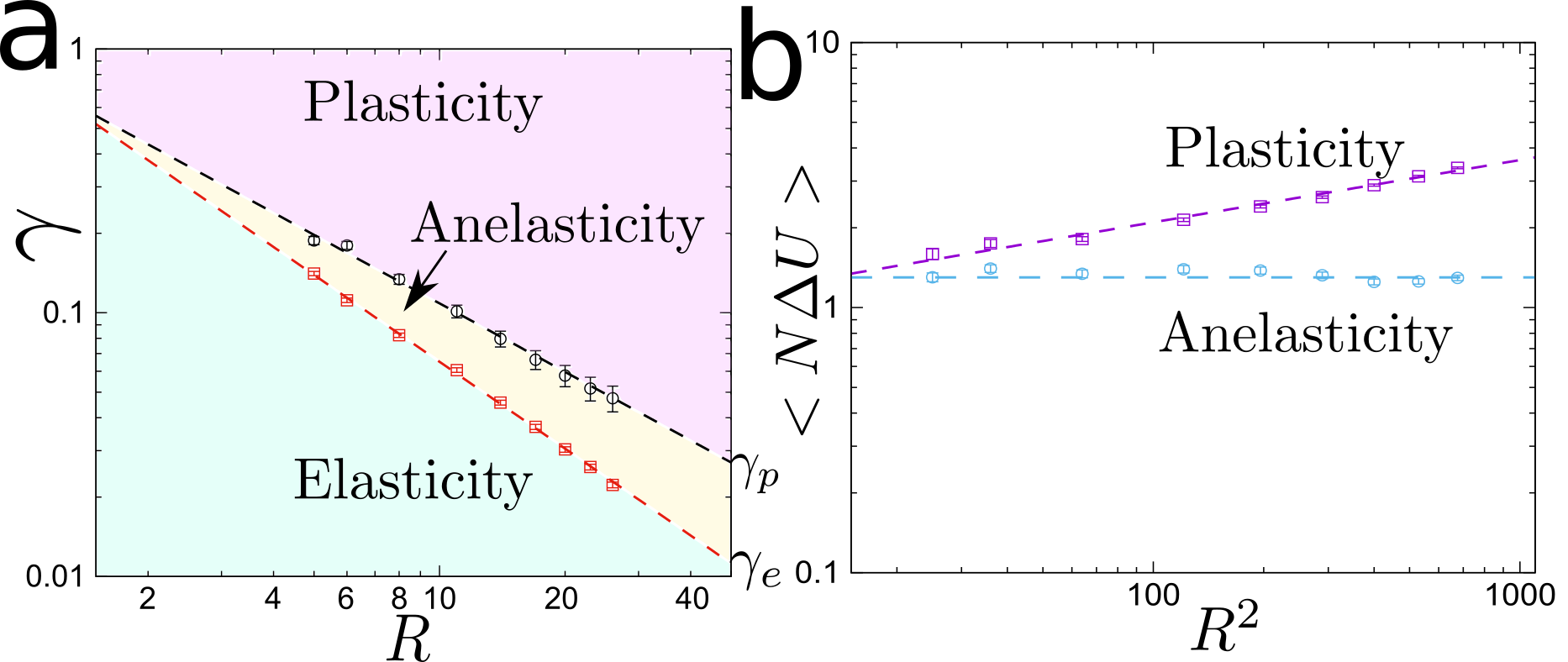}
    \caption{\textbf{The finite size effect of anelasticity and plasticity.}. 
    \textbf{(a)} The strain boundaries of elasticity and anelasticity $\gamma_e$ and of anelasticity and plasticity $\gamma_p$ depend on system size $R$.
    \textbf{(b)} The mean value of avalanche size $N \Delta U$ for  anelasticity and plasticity, respectively, varies with the square of the system size $R^2$.
    } \label{fig:4}
\end{figure}

The anelasticity to plasticity transition is both controlled by system size $R$ and loading strain $\gamma$, and the key parameters for the transition are $\gamma_e$ and $\gamma_p$.
Figure \ref{fig:4} (a) shows the mean value of $\gamma_e$ and $\gamma_p$ decreases with  system size $R$, and the finite size effect can be well depicted by a powerlaw formula $\gamma \sim R^{-\alpha}$, for $\gamma_e \sim R^{-1.09 \pm 0.02}$ and $\gamma_p \sim R^{-0.86 \pm 0.04}$, respectively. 
\BSS{
The discrepancy of exponent between $\gamma_e$ and $\gamma_p$ reveals that, as the system size increase, the anelastic property will be more significant. It will dominate the deformation of the apparent elastic region.
}
Furthermore, for thermodynamic limit $R \to \infty$ , both of elasticity and anelasticity will be disappeared, and the intrinsic mechanical property of amorphous solid is inelasticity, this is consistent with the observation of avalanche statistics in the apparent elastic region\ucite{Shang2019}. 
It confirms that regardless of boundary condition, the nature of amorphous solid is inelastic. 

The finite size exponent can determent the property of avalanche statistic, and the avalanche size is defined as $N \Delta U$. \BSS{The boundary between anelasticity and plasticity is marked by $\gamma_p$. Avalanches during loading between $\gamma_e$ and $\gamma_p$ are anelastic; beyond this range, they are plastic. Figure 4 (b) illustrates avalanche size statistics. Plastic avalanches grow with system size, while anelastic ones stay constant regardless of size.}
It reveals the anelastic avalanche is localized event, which is distinguished with the sub-extensitive nature of plastic avalanche\ucite{PhysRevLett.93.016001,PhysRevE.79.066109}.
The anelastic event can be recognized as the basin hopping within a metabasin based on the view of potential energy landscape\ucite{PhysRevLett.99.135502,PhysRevE.98.033002}, and the accumulation of basin hopping would arouse metabasin hopping, it means the plastic event can be only triggered when the strain is larger than $\gamma_p$.
Therefore, the anelasticity can be identified as the activation of element excitation, such as STZ\ucite{PhysRevE.82.055103} or fluid units\ucite{Wang2018}, and the sub-extensitive plasticity is composed of the element excitation, and entangled with long range elasticity.

\section{Conclusion}

In summary, our study focused on understanding the transition from anelasticity to plasticity in amorphous solids. We employed molecular dynamics simulations using athermal quasistatic shear and frozen matrix methods. By analyzing various parameters, we identified three distinct deformation modes: elasticity, anelasticity, and plasticity.
We found that anelasticity acts as a critical intermediary between elasticity and plasticity, and it precedes the onset of plastic deformation. The transition from anelasticity to plasticity is influenced by both the system size ($R$) and the applied loading strain ($\gamma$). We characterized this transition using key parameters $\gamma_e$ and $\gamma_p$.
 The transition from anelasticity to plasticity occurs when the loading strain exceeds $\gamma_p$. Anelastic events were found to exhibit localized behavior, while plastic events exhibited subextensive behavior.
 
Our findings suggest that plasticity arises from the interaction of a series of anelastic events, with anelastic events acting as potential element excitations in amorphous solids. The transition from anelasticity to plasticity is influenced by system size and loading strain, and the properties of avalanche statistics depend on the system size and the distinction between anelastic and plastic avalanches.



\end{CJK*}  

\addcontentsline{toc}{chapter}{References}
\begin{thebibliography}{99}\footnotesize
\itemsep=-3pt plus.2pt minus.2pt   
  \bibitem{Schuh20074067}
	Schuh C, Hufnagel T and Ramamurty U.
	\href{http://www.sciencedirect.com/science/article/pii/S135964540700122X}
	 {2007 \em Acta Mater.}, \textbf{55}, 4067
  \bibitem{Wang2012a}
	Wang W H.
	\href{http://www.sciencedirect.com/science/article/pii/S0079642511000934}
	 { 2012 \em Prog. Mater. Sci}, \textbf{57}, 487

  \bibitem{hufnagel2016deformation}
	Hufnagel T, Schuh C, and Falk M.
	\href{https://www.sciencedirect.com/science/article/abs/pii/S1359645416300465}
	 {2016 \em Acta Mater.}, \textbf{109}, 375

  \bibitem{Cheng2011379}
	Cheng Y Q and Ma E.
	\href{http://www.sciencedirect.com/science/article/pii/S0079642510000691}
	 {2011 \em Prog. Mater. Sci}, \textbf{56}, 379

  \bibitem{barrat2011heterogeneities}
	Barrat J and Lemaitre A. 2011 \emph{
	Heterogeneities in amorphous systems under shear.},
	(Oxford University Press) pp.~246

	\bibitem{RevModPhys.90.045006}
	  Nicolas A, Ferrero E, Martens K and Barrat J.
	  \href{https://link.aps.org/doi/10.1103/RevModPhys.90.045006}
	   {2018 \em Rev. Mod. Phys.},\textbf{90}, 045006

	\bibitem{argon1979plastic}
	  Argon A.
	  \href{https://www.sciencedirect.com/science/article/abs/pii/0001616079900555}
	   {1979 \em Acta metallurgica}, \textbf{27}, 47

	\bibitem{PhysRevLett.95.095502}
	  Shi Y F and Falk M.
	  \href{http://link.aps.org/doi/10.1103/PhysRevLett.95.095502}
	   {2005 \em Phys. Rev. Lett.}, \textbf{95}, 095502

	\bibitem{schall2007structural}
	  Schall P, Weitz D and Spaepen F.
	  \href{https://www.science.org/doi/10.1126/science.1149308}
	   {2007 \em Science}, \textbf{318}, 1895

	\bibitem{PhysRevE.82.055103}
	  Karmakar S, Lerner E and Procaccia I.
	  \href{http://link.aps.org/doi/10.1103/PhysRevE.82.055103}
	   {2010 \em Phys. Rev. E}, \textbf{82}, 055103

	\bibitem{PhysRevE.96.033002}
	  Lin J and Zheng W.
	  \href{https://link.aps.org/doi/10.1103/PhysRevE.96.033002}
	   {2017 \em Phys. Rev. E}, \textbf{96}, 033002

	\bibitem{PhysRevE.79.066109}
	  Lerner E and Procaccia I.
	  \href{http://link.aps.org/doi/10.1103/PhysRevE.79.066109}
	   {2009 \em Phys. Rev. E}, \textbf{79}, 066109

	\bibitem{PhysRevLett.93.016001}
	  Maloney C and  Lemaitre A.
	  \href{http://link.aps.org/doi/10.1103/PhysRevLett.93.016001}
	   {2004 \em Phys. Rev. Lett.}, \textbf{93}, 016001

	\bibitem{krisponeit2014crossover}
	  Krisponeit J, Pitikaris S, Avila K, K{\"u}chemann S,
	  Kr{\"u}ger A and Samwer K.
	  \href{https://www.nature.com/articles/ncomms4616}
	  {2014 \em Nat. Commun.}, \textbf{5}, 3616

	\bibitem{PhysRevLett.112.155501}
	  Antonaglia J, Wright W, Gu X J, Byer R, Hufnagel T, LeBlanc M, Uhl J and Dahmen K.
	  \href{http://link.aps.org/doi/10.1103/PhysRevLett.112.155501}
	   {2014 \em Phys. Rev. Lett.}, \textbf{112}, 155501

	\bibitem{Lagogianni2018}
	  Lagogianni A, Liu C, Martens K and Samwer K.
	  \href{https://doi.org/10.1140/epjb/e2018-90051-7}
	   {2018 \em Eur. Phys. J. B}, \textbf{91}, 104

	\bibitem{Shang2019}
	  Shang B S, Rottler J, Guan P F and Barrat J.
	  \href{https://link.aps.org/doi/10.1103/PhysRevLett.122.105501}
	   {2019 \em Phys. Rev. Lett.}, \textbf{122}, 105501

	\bibitem{PhysRevMaterials.7.013601}
	  Duan J, Wang Y J, Dai L H and Jiang M Q.
	  \href{https://link.aps.org/doi/10.1103/PhysRevMaterials.7.013601}
	   {2023 \em Phys. Rev. Mater.}, \textbf{7}, 013601

	\bibitem{argon2013strain}
	  Argon A.
	  \href{https://doi.org/10.1080/14786435.2013.798049}
	   {2013 \em Philos. Mag.}, \textbf{93}, 3795

	\bibitem{xu2017strain}
	  Xu B, Falk M, Li J F and Kong L T.
	  \href{https://link.aps.org/doi/10.1103/PhysRevB.95.144201}
	   {2017 \em Phys. Rev. B}, \textbf{95}, 144201

	\bibitem{PhysRevMaterials.4.113609}
	  Richard D, Ozawa M, Patinet S, Stanifer E, Shang B S, Ridout S, Xu B,
	  Zhang G, Morse P, Barrat J, Berthier L, Falk M, Guan P F, Liu A, Martens K, Sastry S, Vandembroucq D, Lerner E and Manning M.
	  \href{https://link.aps.org/doi/10.1103/PhysRevMaterials.4.113609}
	   {2020 \em Phys. Rev. Mater.}, \textbf{4}, 113609

	\bibitem{PhysRevE.102.033006}
	  Ebrahem F, Bamer F and Markert B.
	  \href{https://link.aps.org/doi/10.1103/PhysRevE.102.033006}
	   {2020 \em Phys. Rev. E}, \textbf{102}, 033006

	\bibitem{Regev2015}
	  Regev I, Weber J, Reichhardt C, Dahmen K and Lookman T.
	  \href{https://doi.org/10.1038/ncomms9805}
	   {2015 \em Nat. Commun.}, \textbf{6}, 8805

	\bibitem{PhysRevLett.99.135502}
	  Harmon J, Demetriou M, Johnson W, and Samwer K.
	  \href{http://link.aps.org/doi/10.1103/PhysRevLett.99.135502}
	   {2007 \em Phys. Rev. Lett.}, \textbf{99}, 135502

	\bibitem{Wang2019}
	  Wang W H.
	  \href{https://doi.org/10.1016/j.pmatsci.2019.03.006}
	   {2019 \em Prog. Mater. Sci}, \textbf{106}, 100561

	\bibitem{Costa2022anelastic}
	  Costa M, Londo{\~{n}}o J, Blatter A, Hariharan A,
	  Gebert A,  Carpenter M , and Greer A.
	  \href{https://doi.org/10.1016/j.actamat.2022.118551}
	  {2023 \em Acta Mater.}, \textbf{244}, 118551

	\bibitem{PhysRevB.48.3048}
	  Tomida T and Egami T
	  \href{http://link.aps.org/doi/10.1103/PhysRevB.48.3048}
	   {1993 \em Phys. Rev. B}, \textbf{48}, 3048

	\bibitem{Shang2022cycle}
	  Shang B S, Wang W H, and Guan P F.
	  \href{https://doi.org/10.1016/j.actamat.2021.117557}
	   {2022 \em Acta Mater.}, \textbf{225}, 117557

	\bibitem{PhysRevE.88.062401}
	  Regev I, Lookman T and Reichhardt C.
	  \href{http://link.aps.org/doi/10.1103/PhysRevE.88.062401}
	   {2013 \em Phys. Rev. E}, \textbf{88}, 062401

	\bibitem{PhysRevLett.112.025702}
	  Fiocco D, Foffi G and Sastry S.
	  \href{http://link.aps.org/doi/10.1103/PhysRevLett.112.025702}
	   {2014 \em Phys. Rev. Lett.}, \textbf{112}, 025702

	\bibitem{PhysRevE.97.033001}
	  Barbot A, Lerbinger M, Hernandez-Garcia A, 
	  Garc\'{\i}a-Garc\'{\i}a R, Falk M, Vandembroucq D and
	  Patinet S.
	  \href{https://link.aps.org/doi/10.1103/PhysRevE.97.033001}
	   {2018 \em Phys. Rev. E}, \textbf{97}, 033001

	\bibitem{Thompson2022}
	  Thompson A, Aktulga H, Berger R, Bolintineanu D,
	  Brown W, Crozier P,  in~t~Veld J,
	  Kohlmeyer A, Moore S, Nguyen T, Shan R, Stevens M,
	  Tranchida J, Trott C and Plimpton S.
	  \href{https://doi.org/10.1016/j.cpc.2021.108171}
	   {2022 \em Computer Physics Communications}, \textbf{271}, 108171

	\bibitem{Stukowski2009VisualizationAA}
	  Stukowski A.
	  \href{https://iopscience.iop.org/article/10.1088/0965-0393/18/1/015012/meta}
	   {2009 \em Model. Simul. Mater. Sci. Eng.},
	  \textbf{18}, 015012

	\bibitem{PhysRevLett.117.045501}
	  Patinet S, Vandembroucq D and Falk M.
	  \href{http://link.aps.org/doi/10.1103/PhysRevLett.117.045501}
	   {2016 \em Phys. Rev. Lett.}, \textbf{117}, 045501

	\bibitem{PhysRevE.87.042306}
	  Mizuno H, Mossa S, and Barrat J.
	  \href{http://link.aps.org/doi/10.1103/PhysRevE.87.042306}
	   {2013 \em Phys. Rev. E}, \textbf{87}, 042306

	\bibitem{PhysRevLett.129.195501}
	  Lerbinger M, Barbot A, Vandembroucq D and Patinet S.
	  \href{https://link.aps.org/doi/10.1103/PhysRevLett.129.195501}
	   {2022 \em Phys. Rev. Lett.}, \textbf{129}, 195501

	\bibitem{FernndezCastellanos2021}
	  Castellanos D, Roux S and Patinet S.
	  \href{https://doi.org/10.5802/crphys.48}
	   {2021 \em Comptes Rendus. Physique}, \textbf{22}, 135

	\bibitem{Castellanos2022}
	  Castellanos D, Roux S and Patinet S.
	  \href{https://doi.org/10.1016/j.actamat.2022.118405}
	   {2022 \em Acta Mater.}, \textbf{241}, 118405

	\bibitem{lemaitre2006sum}
	  Lemaitre A and Maloney C.
	  \href{https://link.springer.com/article/10.1007/s10955-005-9015-5}
	   {2006 \em J. Stat. Phys.}, \textbf{123}, 415

	\bibitem{PhysRevE.98.033002}
	  Blank-Burian M and Heuer A.
	  \href{https://link.aps.org/doi/10.1103/PhysRevE.98.033002}
	   {2018 \em Phys. Rev. E}, \textbf{98}, 033002

	\bibitem{Wang2018}
	  Wang Z and Wang W H.
	  \href{https://doi.org/10.1093/nsr/nwy084}
	   {2018 \em Natl. Sci. Rev.}, \textbf{6}, 304

  \end{thebibliography}
\end{document}